\documentclass[12pt]{JHEP3}
\usepackage{graphicx}
\usepackage{epsfig}
\usepackage{amssymb}
\usepackage{bm}
\usepackage{amsmath}
\newcommand{\bea}{\begin{eqnarray}}
\newcommand{\eea}{\end{eqnarray}}

\newcommand{\bra}{\langle}
\newcommand{\ket}{\rangle}

\makeatletter
\@addtoreset{equation}{section}
\makeatother

\title{
Magnetic instability in AdS/CFT : Schwinger effect and Euler-Heisenberg Lagrangian
of Supersymmetric QCD
}
\author{
Koji Hashimoto$^{1,2,*}$, 
Takashi Oka$^{3,\#}$ and Akihiko Sonoda$^{1,\dagger}$
\\

$^1$
{\it Department of Physics, Osaka University,
Toyonaka, Osaka 560-0043, Japan}\\
$^2$
{\it Mathematical Physics Lab., RIKEN Nishina Center,
Saitama 351-0198, Japan}\\
$^3$
{\it Department of Applied Physics, University of Tokyo, 
Tokyo 113-8656, Japan}\\
E-mail: $^*$ \email{koji(at)phys.sci.osaka-u.ac.jp}\\ 
E-mail: $^\#$ \email{oka(at)ap.t.u-tokyo.ac.jp}\\
E-mail: $^\dagger$ \email{sonoda(at)het.phys.sci.osaka-u.ac.jp}\\ 
}

\abstract{
To reveal the Schwinger effect for quarks, i.e., 
pair creation process of quarks and antiquarks,
we derive the vacuum decay rate 
at strong coupling using AdS/CFT correspondence.
Magnetic fields, in addition to the electric field responsible for the pair creation,
causes prominent effects on the rate, and is important also in experiments such 
as RHIC/LHC heavy ion collisions.
In this paper, through the gravity dual we obtain the full
Euler-Heisenberg Lagrangian of ${\cal N}=2$ supersymmetric QCD
and study the Schwinger mechanism with not only a constant electric field but also a constant
magnetic field as external fields.
We determine the quark mass and temperature dependence of the Lagrangian. 
In sharp contrast with the zero magnetic field case, 
we find that the imaginary part, and thus the vacuum decay rate, diverges in the massless
zero-temperature limit.
This may be related to a strong instability of the QCD vacuum 
in strong magnetic fields.
The real part of the Lagrangian serves as a generating function
for non-linear electro-magnetic responses,
and is found such that the Cotton-Mouton effect 
vanishes. Interestingly,
our results of the Schwinger / Cotton-Mouton effects coincide precisely with those of ${\cal N}=2$ supersymmetric QED.
}

\preprint{
{\normalsize OU-HET-811} \\
{\normalsize RIKEN-MP-86}
}


\begin{document}
\setcounter{page}{1}

\section{Introduction}

The renowned Schwinger effect, 
creation process of electron-positron pairs 
in strong electric fields, is a big challenge 
in the field of non-linear quantum field theory. Although the
Schwinger limit $E \sim m_e^2$ has not been reached by the direct experiments such as strong lasers, 
similar effective setups in materials are actively investigated. Theoretical foundation of the Schwinger effect  \cite{Schwinger:1951}\footnote{For reviews, see \cite{Dunne:2004nc,Dittrich}.} was to evaluate the imaginary part of the effective action
of QED under a constant electromagnetic field, the Euler-Heisenberg Lagrangian
\cite{Heisenberg:1935qt} which dates back to 1936. 
The Euler-Heisenberg Lagrangian
is a generating function 
of nonlinear electromagnetic responses of the vacuum.
In its expression, the electric field couples to the magnetic field in
a complicated and nonlinear manner, and the total effective Lagrangian is a starting point 
in the research of strong fields in QED, including the non-perturbative Schwinger effect.

In \cite{Hashimoto:2013mua}, two of the present 
authors derived an Euler-Heisenberg Lagrangian for a supersymmetric QCD in the strong coupling limit, by using the AdS/CFT correspondence \cite{Maldacena:1997re,Gubser:1998bc,Witten:1998qj}. Since the quarks have electric charges, once a strong electric field (not a color electric field\footnote{For the color electric field and its Schwinger effect, see for example
\cite{Yildiz:1979vv,Ambjorn:1982bp,Tanji:2008ku,Tanji:2011di}. We use the AdS/CFT correspondence, so it is challenging to find how to treat color electromagnetic fields.} 
but the Maxwell electric field) is applied, a quark antiquark pair is created. The nontrivial part is the gluon interaction at strong coupling in QCD. The quarks are confined, and between the quark and the antiquark a confining force (a QCD string) is present to bind them. If the electric field is strong enough, the quarks are liberated. This truly nonperturbative process
is of importance, not only because it can be a realistic phenomenon occurring in the universe, but also because it may be a touchstone for understanding the quark confinement. 

There are at least two cases in which the QCD Schwinger effect may play an important role: First, the heavy ion collision experiment, and second, magnetars (neutron stars with 
a very strong magnetic field). In the heavy ion collisions, very strong electric fields are generated by the 
the electric current induced by heavy ions passing by each other. Since the
magnetic field is time dependent, there appears strong electric field as well \cite{Kharzeev:2007jp,Skokov:2009qp,Voronyuk:2011jd,Bzdak:2011yy,Deng:2012pc}, and 
it may be related to the formation of the quark gluon plasma. On the other hand, magnetars are known to be the most dense place in the universe, and the strong magnetic field accompanied by 
some electric field can occur and affect the core structure of the stars, possibly having a quark phase inside. In these examples, the understanding of QCD
and Schwinger effect in strong electric and magnetic fields
can be tested by experiments/observations
and serves as a playground at which we can test our 
theoretical knowledge on strongly coupled quantum field theories.

Via the AdS/CFT correspondence, the Schwinger effect of ${\cal N}=4$ supersymmetric Yang-Mills theory at strong coupling was calculated in \cite{Semenoff:2011ng}, where an 
explicit saddle point 
analysis of a quark antiquark pair (a one-instanton calculation) was made. Based on
\cite{Semenoff:2011ng},
varieties of calculations of the one-instanton amplitude were reported 
\cite{Ambjorn:2011wz,Bolognesi:2012gr,Sato:2013pxa,Sato:2013iua,Sato:2013dwa,
Sato:2013hyw,Kawai:2013xya,Sakaguchi:2014gpa}. In these papers, a string worldsheet shape in the AdS space was evaluated to calculate the 
single pair creation\footnote{For brane creation approach, see \cite{Gorsky:2001up}. For an application to EPR, see \cite{Sonner:2013mba,Chernicoff:2013iga}.}. On the other hand, two of the present authors took a different approach
\cite{Hashimoto:2013mua} at which a D-brane action in AdS was found to be directly equivalent to the 
effective Lagrangian (Euler-Heisenberg Lagrangian) including the imaginary part 
agreeing with instanton summation of the Schwinger effect, for a large electric field. It describes also the real
part, and the case where the worldsheet approach breaks down at the large electric field.

The result of \cite{Hashimoto:2013mua} is summarized as follows; The Euler-Heisenberg Lagrangian of strongly coupled ${\cal N}=2$ supersymmetric QCD at large $N_c$ limit
was calculated in 
the presence of a constant electric field using the AdS/CFT correspondence. Its imaginary part explicitly evaluated is found to agree with large electric field expansion of the Schwinger effect of ${\cal N}=2$ supersymmetric QED (once the QCD string tension is replaced by the electron mass). However, there, only the electric field was considered.
In this paper, we include the full dependence of the magnetic field, which is 
important as is obvious from the physical situations explained above. 

Here we summarize the finding of the present paper: 
\begin{itemize}
\item We obtain the Euler-Heisenberg Lagrangian of the $\mathcal{N}=2$ supersymmetric QCD in a constant electromagnetic field, at strong coupling and large $N_c$ limit. 
\item We evaluate the imaginary part of the Euler-Heisenberg Lagrangian, and find that the rate of the quark antiquark creation diverges at zero temperature for massless quarks.
\item The divergent rate can be regularized, i.e., the vacuum is unstable but
the lifetime becomes finite, by either introducing 
finite temperature or a quark mass.
\item We compute the real part of the Euler-Heisenberg Lagrangian, and show the disappearance of Cotton-Mouton effect in an expansion with the electromagnetic field.
\item 
The imaginary part of the Euler-Heisenberg Lagrangian for a small quark mass
is shown to coincide
with that of ${\cal N}=2$ supersymmetric QED, at the leading order in electron mass.
The agreement is found also for the real parts responsible for the Cotton-Mouton effect.
\end{itemize}
\hspace{3mm}The organization of this paper is as follows. In section 2, using the AdS/CFT correspondence, we obtain the effective action (Euler-Heisenberg Lagrangian) 
of $\mathcal{N}=2$\hspace{1mm} massless supersymmetric QCD (SQCD) in a constant electromagnetic field. It is given as a function of
the field strengths $\vec{E}$ and  $\vec{B}$, the charge density $d$ and the current $j$.
For this we use an extension of the dictionary of the AdS/CFT correspondence and the equations of motion.
Then we compute the imaginary part of the Euler-Heisenberg Lagrangian 
for $d=j=0$. 
We find that the rate of the quark antiquark creation is divergent, and
has a regularized form $\log T$ for a small temperature $T$. 
In section 3, we obtain the Euler-Heisenberg Lagrangian in gapped system when the quark mass is nonzero, and see that the divergent 
imaginary part is regularized by the small quark mass. We 
evaluate the real part to study the non-linear electro-magnetic
response of the vacuum, and at the third order,
we find that the Cotton-Mouton effect disappears. 
We compare these results with those of weakly coupled supersymmetric QED and 
find a quantitative agreement, unexpectedly. 
Finally in section 4, we summarize our results.

\vspace*{20mm}
\section{Euler-Heisenberg Lagrangian with electromagnetic field in massless SQCD}

It is a challenging problem to evaluate Schwinger effect in strongly coupled gauge theories.
The AdS/CFT correspondence is a universal tool for analyzing such theories.
We can consider a classical gravity which is dual to the strongly coupled gauge theories 
using the AdS/CFT correspondence. 
In this section, we obtain the Euler-Heisenberg Lagrangian, which is the effective Lagrangian
in an external electromagnetic field, for a massless supersymmetric QCD, and evaluate the Lagrangian.
First, in section 2.1, we calculate the full Euler-Heisenberg Lagrangian as a function of the 
electromagnetic fields $\vec{E}$ and $\vec{B}$, the quark number density $d$ and
the electric current $j$. Then in section 2.2, we evaluate the imaginary part and see the
divergence due to the presence of the magnetic field. We see that the divergence is regularized by a
temperature.


\subsection{Full Euler-Heisenberg Lagrangian of strongly coupled ${\cal N}=2$ SQCD}

Consider the supersymmetric gauge theory on the boundary of an AdS space, 
which is $\mathcal{N}=4$ supersymmetric Yang-Mills theory with a 
$\mathcal{N}=2$ hypermultiplet with $SU(N_{c})$ gauge group. 
The configuration of D-branes 
realizing the gauge theory is a D3-D7 brane system \cite{Karch}. Taking a gravity dual,
we use an AdS black hole metric as a background metric, in order to see the relation between the rate of quark antiquark creation and a temperature.
The probe D7-brane in the AdS space has been well-studied to look at 
an electric conductivity, that is, an electric current determined by the external electric field \cite{Karch2,Karch3,Erdmenger}.\footnote{See also \cite{Hartnoll:2007ip,Albash:2007bk,Albash:2007bq,Johnson:2008vna,Bergman:2008sg,Kim:2008zn,Evans:2011mu,Alam:2012fw} for related references.} 
The effective action (the D7-brane action in the AdS) is put to be real to determine the conductivity. Here on the other hand, we are interested in the instability caused by the external
electromagnetic field, so the current $j$ is put to zero (or takes some arbitrary value),
giving an imaginary part in the effective action: this is how we obtain the Euler-Heisenberg Lagrangian \cite{Hashimoto:2013mua}.

The AdS black hole metric is the following,
\begin{align}
ds^{2} = \frac{R^{2}}{z^{2}}\left[-\left(1-\frac{z^{4}}{z^{4}_{H}}\right)dt^{2} + \left(1-\frac{z^{4}}{z^{4}_{H}}\right)^{-1}dz^{2} + d\Vec{x}^{2}\right] + R^{2} d\Omega^{2}_{5},
\end{align}
where the coordinate $z$ is the AdS radial direction, and $z=0$ corresponds to the boundary of the AdS space, and $z=z_{H}$ is the horizon of the black hole. $d\vec{x}^{2}$ is defined by $d\vec{x}^{2}=dx_{1}^{2}+dx_{2}^{2}+dx_{3}^{2}$. $R$ is the radius of the AdS space.  These parameters are given by the following relations between that of the gauge side and that of the gravity side:
\begin{align}
z_{H} = \frac{1}{\pi{T}}\hspace{13mm}R^{4} = 2\lambda \alpha'^{2},
\end{align}
where T is a temperature, and $\lambda \equiv N_c g_{\mathrm{QCD}}^2$ is a 't Hooft coupling of the SQCD, and $\alpha'$ is defined by $\alpha'=l^{2}_{s}$.

In the D3-D7 system,
the D7-brane has the degree of freedom of the quark in the fundamental representation of the color $SU(N_{c})$ gauge group by coupling to the D3-branes. Going to the gravity dual, 
the action in the gravity side is a D7-brane action in the AdS space. 
The D7-brane action is the following:
\begin{align}
S_{D7} = - \mu_{7}\int{dtd^{3}\vec{x}dzd^{3}\Omega_{3}}\sqrt{-\mathrm{det}\left[P[g]_{ab} + 2\pi\alpha'F_{ab}\right]},
\end{align}
where we need not consider the scalar fields on the D7-brane, since we are working for the massless SQCD. 

The Euler-Heisenberg effective action is defined by
\begin{eqnarray}
\mathcal{L}=-i\ln\bra e^{-i\int A_\mu^{\rm ext} j^\mu}\ket_0,
\label{eq:EHdef}
\end{eqnarray}
which is a function of the external electric and 
magnetic fields represented by $A_\mu^{\rm ext}$.
$j^\mu$ is the $U(1)$ current operator corresponding to the baryon charge, 
and the expectation value $\bra \;\ket_0$ is taken with respect to the
``false vacuum", i.e., the vacuum without the field which is now unstable.
If the expectation value in (\ref{eq:EHdef}) was taken 
with the true vacuum, 
the standard AdS/CFT dictionary \cite{Gubser:1998bc,Witten:1998qj}
states that the effective action is given by the 
D7-brane action evaluated with the 
reality condition for the action \cite{Karch2,Karch3,Erdmenger} and with the
solution of the equation of motion. 
In ref.~\cite{Hashimoto:2013mua}, two of the present authors proposed that 
the effective action (\ref{eq:EHdef}) is given as the
D7-brane action evaluated with the false-vacuum solution,
i.e., the solution of the equation of motion without the electromagnetic fields.
Substituting the AdS black hole metric for the D7-brane action, the effective action becomes 
\begin{align}
\mathcal{L} = -2\pi^{2}\mu_{7}\int{dz}\frac{R^{8}}{z^{5}}\sqrt{\xi},
\end{align}
where $d\Omega$-integral is $\mathrm{Vol}(S^{3})=2\pi^{2}$. The factor 
$\mu_{7}$ is the D7-brane tension, given by $\mu_{7}\equiv 1/((2\pi)^{7}g_{s}\alpha'^{4})$. 
The string coupling constant $g_s$ is related to the gauge coupling constant of SQCD as
$2\pi g_s = g_{\rm QCD}^2$.
Without losing generality, we can choose the direction of the electromagnetic fields.
Using the rotation symmetry, we fix the electric field to the $x_1$ direction. 
The magnetic fields  are generic 
in $x_{1},x_{2},x_{3}$ directions. Then the $\xi$ in the action is defined by the following:
\begin{align}
\xi \equiv 1 - \frac{(2\pi\alpha')^{2}z^{4}}{R^{4}}\left[F_{0z}^{2} + F_{01}^{2}h(z)^{-1} - F_{1z}^{2}h(z) - F_{12}^{2} - F_{23}^{2} - F_{13}^{2}\right] \notag \\
- \frac{(2\pi\alpha')^{4}z^{8}}{R^{8}}[F_{23}^{2}\{F_{01}^{2}h(z)^{-1} - F_{1z}^{2}h(z)\} + F_{0z}^{2}\left\{F_{12}^{2} + F_{23}^{2} + F_{13}^{2}\right\}].
\label{xiEB}
\end{align}
The function $h(z)$ is defined as $h(z)=1 - z^{4}/z_{H}^{4}$.

Consider rewriting the effective Lagrangian (\ref{xiEB})
in order to see the dependence on the charge density $d$ and the current $j$. 
We derive the equations of motion from this action. Since we are interested in homogeneous phases, we simply put $\partial_{i}=0\hspace{1mm}(i=1,2,3)$. 
Then the equations of motion are the following: 
\begin{align}
\partial_{z}\left[\frac{F_{0z}}{z\sqrt{\xi}} + \frac{(2\pi\alpha')^{2}z^{3}}{R^{4}\sqrt{\xi}}F_{0z}(F_{12}^{2} + F_{23}^{2} + F_{13}^{2})\right] = 0,
\end{align}
\begin{align}
\partial_{0}\left[\frac{F_{0z}}{z\sqrt{\xi}} + \frac{(2\pi\alpha')^{2}z^{3}}{R^{4}\sqrt{\xi}}F_{0z}(F_{12}^{2} + F_{23}^{2} + F_{13}^{2})\right] = 0,
\end{align}
\begin{align}
\partial_{0}\left[\frac{F_{01}}{z\sqrt{\xi}}h(z)^{-1} \!+\! \frac{(2\pi\alpha')^{2}z^{3}}{R^{4}\sqrt{\xi}}F_{01}F_{23}^{2}h(z)^{-1}\right] \!+ \!\partial_{z}\left[\frac{F_{1z}}{z\sqrt{\xi}}h(z) + \frac{(2\pi\alpha')^{2}z^{3}}{R^{4}\sqrt{\xi}}F_{1z}F_{23}^{2}h(z)\right]= 0.
\end{align}
In particular, the equations of motions in the case of time-independent 
field configurations are 
\begin{align}
\partial_{z}\left[\frac{F_{0z}}{z\sqrt{\xi}} + \frac{(2\pi\alpha')^{2}z^{3}}{R^{4}\sqrt{\xi}}F_{0z}(F_{12}^{2} + F_{23}^{2} + F_{13}^{2})\right] = 0,
\end{align}
\begin{align}
\partial_{z}\left[\frac{F_{1z}}{z\sqrt{\xi}}h(z) + \frac{(2\pi\alpha')^{2}z^{3}}{R^{4}\sqrt{\xi}}F_{1z}F_{23}^{2}h(z)\right]= 0.
\end{align}

Next let us evaluate the charge density $d$ and the current $j$ in the gauge side.  Using the dictionary of the AdS/CFT correspondence, they are respectively
\begin{align}
d = \frac{2\pi\alpha'F_{0z}}{z\sqrt{\xi}} + \frac{(2\pi\alpha')^{3}z^{3}}{R^{4}\sqrt{\xi}}F_{0z}(F_{12}^{2} + F_{23}^{2} + F_{13}^{2}),
\end{align}
\begin{align}
j = \frac{2\pi\alpha'F_{1z}}{z\sqrt{\xi}}h(z) + \frac{(2\pi\alpha')^{3}z^{3}}{R^{4}\sqrt{\xi}}F_{1z}F_{23}^{2}h(z).
\end{align}
These charge density and current are substituted into $\xi$, to find
\begin{align}
\xi = \frac{1 - \frac{(2\pi\alpha')^{2}z^{4}}{R^{4}}(E_{1}^{2}h(z)^{-1} - \vec{B}^{2}) - \frac{(2\pi\alpha')^{4}z^{8}}{R^{8}}(E_{1}B_{1})^{2}h(z)^{-1}}{1 + \frac{z^{6}d^{2}}{R^{4}\left(1 + \frac{(2\pi\alpha')^{2}z^{4}\vec{B}^{2}}{R^{4}}\right)} - \frac{z^{6}j^{2}h(z)^{-1}}{R^{4}\left(1 + \frac{(2\pi\alpha')^{2}z^{4}B_{1}^{2}}{R^{4}}\right)}},
\end{align}
where $F_{01}\equiv E_1$ is a constant electric field, and $F_{12}\equiv B_{3}$, $F_{23}\equiv B_{1}$, $F_{31}\equiv B_{2}$ are constant magnetic fields. 
Using this $\xi$, 
the effective Lagrangian with the constant electromagnetic fields is as follows:
\begin{align}
\mathcal{L} = -2\pi^{2}\mu_{7}\int_0^{z_H}{dz}\frac{R^{8}}{z^{5}}\sqrt{\frac{1 - \frac{(2\pi\alpha')^{2}z^{4}}{R^{4}}(E_{1}^{2}h(z)^{-1} - \vec{B}^{2}) - \frac{(2\pi\alpha')^{4}z^{8}}{R^{8}}(E_{1}B_{1})^{2}h(z)^{-1}}{1 + \frac{z^{6}d^{2}}{R^{4}\left(1 + \frac{(2\pi\alpha')^{2}z^{4}\vec{B}^{2}}{R^{4}}\right)} - \frac{z^{6}j^{2}h(z)^{-1}}{R^{4}\left(1 + \frac{(2\pi\alpha')^{2}z^{4}B_{1}^{2}}{R^{4}}\right)}}}.
\label{EHL}
\end{align}
When $j=0$, the solution corresponds to the false vacuum,
and (\ref{EHL}) gives the Euler-Heisenberg effective action. 
This is our result from the AdS/CFT correspondence, and the basis for the following analyses. 
The imaginary part of the effective action
gives half the inverse life time of the false vacuum. 

When $j=0$, using the spatial rotation symmetry, we can recover the full ${\vec{E}}$ and $\vec{B}$ dependence.
The Euler-Heisenberg Lagrangian for a generic constant electromagnetic field, at a finite temperature is
\begin{eqnarray}
{\cal L} = -2\pi^2 \mu_7
\int_0^{z_H} dz
\frac{R^8}{z^5}\sqrt{
\frac{1 + \beta(z) \vec{B}^2 - \frac{\beta(z)}{h(z)}\vec{E}^2 - \frac{\beta(z)^2}{h(z)}
\left(\vec{E}\cdot\vec{B}\right)^2}{1+ 
\frac{z^2}{(2\pi\alpha')^2} \frac{\beta(z)}{1+\beta(z) \vec{B}^2} d^2}
}
\end{eqnarray}
and $\beta(z) \equiv (2\pi\alpha')^2 z^4/R^4$. In particular, for the vanishing density $d=0$, the Euler-Heisenberg Lagrangian is simplified as
\begin{eqnarray}
{\cal L} = -2\pi^2 \mu_7
\int_0^{z_H} dz
\frac{R^8}{z^5}\sqrt{
1 + \beta(z) \vec{B}^2 - \frac{\beta(z)}{h(z)}\vec{E}^2 - \frac{\beta(z)^2}{h(z)}
\left(\vec{E}\cdot\vec{B}\right)^2
}
\label{masslessxi}
\end{eqnarray}

In the language of the massless ${\cal N}=2$ SQCD, this Euler-Heisenberg Lagrangian 
(at a finite temperature and with $d=j=0$) is written as
\begin{eqnarray}
{\cal L} = -\frac{N_c \lambda}{2^3\pi^4} 
\int_0^{1/(\pi T)} 
\frac{dz}{z^5}\sqrt{
1 + \beta(z) \vec{B}^2 - \frac{\beta(z)}{h(z)}\vec{E}^2 - \frac{\beta(z)^2}{h(z)}
\left(\vec{E}\cdot\vec{B}\right)^2
}
\end{eqnarray}
where $\beta(z) =(2\pi^2/\lambda)z^4$ and $h(z)=1-(\pi T)^4 z^4$.


\subsection{Magnetic instability and imaginary part of Lagrangian}

In this subsection, we evaluate the imaginary part of the effective Lagrangian, and study the vacuum instability against not only the electric field but also the magnetic field. 
First, the imaginary part at zero temperature $T=0$ and zero quark density
diverges: The vacuum is not protected by a gap and thus extremely unstable. 
In a finite temperature case, the divergence is suppressed.
In fact, assuming that the temperature provides a thermal mass for the quarks, the divergence of the imaginary part 
coincides with the result of a massive SQCD, and further with a supersymmetric 
QED, as we shall see in the next section.

In the previous subsection, we obtained  the effective Lagrangian (\ref{EHL}) with not only the constant electric field but also the constant magnetic field in the massless system. 
For simplicity, consider the case when the magnetic field is parallel to the electric field 
($E_{1}$ and $B_{1}$ are nonzero).
Then the effective Lagrangian is given by
\begin{align}
\mathcal{L} = -2\pi^{2}\mu_{7}\int_{0}^{z_{H}}{dz}\frac{R^{8}}{z^{5}}\sqrt{1 - \frac{(2\pi\alpha')^{2}z^{4}}{R^{4}}(E_{1}^{2}h(z)^{-1} - {B}_{1}^{2}) - \frac{(2\pi\alpha')^{4}z^{8}}{R^{8}}(E_{1}B_{1})^{2}h(z)^{-1}}.
\end{align}
We evaluate the imaginary part of the effective Lagrangian to derive the rate of the quark antiquark creation. 

Consider the  zero-temperature case, i.e., $z_{H}\to\infty$. Then the function $h(z)$ approaches unity. The $z$-integral of the imaginary part of the effective Lagrangian is dominated by the third term in the square root of the Lagrangian. Thus, this $z$-integral has a logarithmic divergence. 
Thus, in the presence of the magnetic field in addition to the electric field,
the vacuum decay rate diverges for massless SQCD at strong coupling  and at zero temperature. 
This is in sharp contrast with the zero magnetic field case
in which $\mbox{Im}\mathcal{L}=\frac{N_c}{32\pi}E^2$ is obtained
\cite{Hashimoto:2013mua}.
In free Dirac systems, it is known that the 
divergence of the Euler-Heisenberg effective Lagrangian
depends on dimensionality. 
In the pure electric field case (no magnetic field), for spatial dimension larger than two, the decay rate is finite, while 
for a (1+1)-dimensional system a divergence takes place. 
Our finding can be understood as an effective dimension reduction
by the magnetic field.
In a finite magnetic field, Landau levels
are formed and the dispersion becomes flat in the two directions 
perpendicular to the field. Starting from three spatial dimensions,
the magnetic field reduces the effective dimension
to one. This may explain the divergence we obtain,
although, it is unclear if the argument holds for 
a strongly interacting model.

In a finite temperature system, the divergence of the decay rate is suppressed.
In order to evaluate the imaginary part of effective Lagrangian, we change the variable $z$ of this integral to $y$ defined by $y\equiv{z}/z_{H}$,
\begin{align}
\mathcal{L} = -2\pi^{2}\mu_{7}(2\pi\alpha')^{2}R^{4}\chi\int^{1}_{0}\frac{dy}{y^{5}}\sqrt{1 - \frac{y^{4}}{\chi}(E_{1}^{2}(1-y^{4})^{-1} - {B}_{1}^{2}) - \frac{y^{8}}{\chi^{2}}(E_{1}B_{1})^{2}(1-y^{4})^{-1}}\, ,
\label{chiex}
\end{align}
where $\chi$ is defined as $\chi\equiv{R^{4}/(2\pi\alpha')^{2}z_{H}^{4}}$. 
As mentioned above, we found that the imaginary part of the Lagrangian diverges in the 
limit $T\rightarrow 0$.
In order to see the dependence on $\chi$ in the square root, we
further change the variable $y$ to $Y\equiv\chi^{-\frac{1}{4}}y$,
\begin{align}
\mathcal{L} = -2\pi^{2}\mu_{7}(2\pi\alpha')^{2}R^{4}\int^{\chi^{-\frac{1}{4}}}_{0}\!\!\!dY\;
\frac{\sqrt{1 - (\chi + E_{1}^{2} - B_{1}^{2})Y^{4} - \chi{B}_{1}^{2}Y^{8} - (E_{1}B_{1})^{2}Y^{8}}
}{Y^{5}\sqrt{1-\chi{Y}^{4}}}\, .
\end{align}
Let us look for the value of $Y$ at which the integrand turns from real to imaginary. 
Since $\chi$ is small, we can ignore ${\cal O}(\chi)$ term in the numerator, to find the value as $Y=1/\sqrt{E_{1}}$.
So the imaginary part of ${\cal L}$ is from the integral over the region $1/\sqrt{E_{1}} < Y < \chi^{-\frac{1}{4}}$.
At the integration, the forth term in the square root of the numerator in the integrand becomes dominant 
for small $\chi$, hence the imaginary part of the Lagrangian is approximately given by
\begin{align}
\mathrm{Im}\; \mathcal{L} \sim 2\pi^{2}\mu_{7}(2\pi\alpha')^{2}R^{4}(E_{1}B_{1})\int^{\chi^{-\frac{1}{4}}}_{1/\sqrt{E_{1}}}\frac{dY}{Y\sqrt{1 - \chi{Y^{4}}}}\, .
\end{align}
Performing the integral for small $\chi$ leads to 
\begin{align}
\mathrm{Im} \; \mathcal{L}\sim \frac{N_{c}}{4\pi^{2}}E_{1}B_{1}\mathrm{log}\hspace{1mm}\frac{1}{T}.
\label{resultdiv}
\end{align}
where we find a logarithmic dependence on the temperature.\footnote{
Divergences appearing in the DBI action and its relation to the validity of the DBI action
in the probe limit, see \cite{Bigazzi:2013jqa}.}


It is straightforward to obtain the Euler-Heisenberg Lagrangian for a generic constant electromagnetic field in finite temperature. 
For small $T$, we obtain the following dominant term in the divergence,\begin{eqnarray}
{\rm Im}\; {\cal L}_{T\neq 0}
= \frac{N_c}{4\pi^2}
\left|\vec{E}\cdot\vec{B}\right| \log \frac{b(E,B)}{T} + {\cal O}(T^0)\, .
\label{LT}
\end{eqnarray}
Here the constant $b$ is determined as
\begin{eqnarray}
b(E,B) \equiv \left[
\frac12 \left(
\vec{E}^2-\vec{B}^2+\sqrt{\left(\vec{E}^2-\vec{B}^2\right)^2
+ 4\left(\vec{E}\cdot \vec{B}\right)^2}
\right)
\right]^{1/4} \, .
\label{defb}
\end{eqnarray}
In addition, in a finite density system, 
we find a density dependence ($d\neq 0$ but at $T=0$),
\begin{eqnarray}
{\rm Im}\; {\cal L}_{d\neq 0}
= \frac{N_c}{4\pi^2}
\left|\vec{E}\cdot\vec{B}\right| \log \frac{1}{d} + \cdots\, .
\end{eqnarray}
This logarithmic dependence is quite similar to the 
finite temperature case (\ref{LT}). 
In fact, we see in the following that this form is quite common
and has a physical interpretation, see (\ref{regdiv}).

\vspace*{18mm}
\section{Euler-Heisenberg Lagrangian with electromagnetic field in massive SQCD}

In this section, we evaluate the Euler-Heisenberg Lagrangian for the strongly coupled 
${\cal N}=2$ SQCD with a quark mass in constant electromagnetic fields.
We evaluate the imaginary part of the Euler-Heisenberg Lagrangian. 
The leading term coincides with that of a weakly coupled supersymmetric QED. 

\subsection{Critical field}

In this section, for simplicity, we consider $T=0$ and $d=0$. 
First, let us take a 
D7-brane configuration of the SQCD with vanishing electromagnetic fields. The induced metric on the D7-brane is \cite{Karch}
\begin{eqnarray}
ds^2 = \frac{R^2}{z^2} \tilde{h}(z) \; (dx^\mu)^2 + \frac{R^2}{z^2 \tilde{h}(z)}
\left(dz^2 + d\Omega_3^2
\right)\, ,
\end{eqnarray}
where $\mu = 0,1,2,3$ and $d\Omega_3^2$ is the metric of a unit 3-sphere, and
\begin{eqnarray}
\tilde{h}(z) \equiv1 + \frac{z^2\eta^2}{R^2}\, .
\end{eqnarray}
The constant $\eta$ specifies the location of the D7-brane, which
is physically related to the quark mass by
\begin{eqnarray}
\frac{\eta}{2\pi\alpha'} = m_q \, . 
\end{eqnarray}
Turning on the constant electromagnetic fields on the D7-brane, we obtain the
Euler-Heisenberg Lagrangian for the strongly coupled ${\cal N}=2$ SQCD
\begin{eqnarray}
{\cal L} = -2\pi^2 \mu_7
\int_0^\infty \!\! dz \; \frac{R^8}{z^5}\sqrt{\xi} 
\label{xires}
\end{eqnarray}
with
\begin{eqnarray}
\xi \equiv 1 + \beta(z) \tilde{h}(z)^{-2} \left(\vec{B}^2-\vec{E}^2\right)
- \beta(z)^2 \tilde{h}(z)^{-4} \left(\vec{E}\cdot \vec{B}\right)^2 \, .
\end{eqnarray}
The $\vec{B}=0$ result agrees with \cite{Hashimoto:2013mua},
and when $\eta = 0$, it reproduces our massless SQCD result
(\ref{masslessxi}) at $T=0$.

We can evaluate the critical electric field above which the effective Lagrangian acquires
an imaginary part. Solving $\xi=0$ for $z$ results in and equation
\begin{eqnarray}
\tilde{h}(z)^2 = \beta(z) b^4
\end{eqnarray}
where the constant $b(E,B)$ is defined in (\ref{defb}). This equation is simplified as
\begin{eqnarray}
1 = \left(
\frac{2\pi\alpha'}{R^2} b^2 - \frac{\eta^2}{R^4}
\right)z^2 \, .
\label{z0def}
\end{eqnarray}
In order for this to have a solution, we need
\begin{eqnarray}
b(E,B) > \frac{\eta}{\sqrt{2\pi\alpha'}R} = \left(\frac{2\pi^2}{\lambda}\right)^{1/4} m_q \, .
\label{limitE}
\end{eqnarray}
This is the condition for having an imaginary part in the Euler-Heisenberg
Lagrangian, a signal for vacuum instability.
Without the magnetic field ($\vec{B}=0$), this condition 
(\ref{limitE}) reduces to 
the critical electric field found in \cite{Hashimoto:2013mua},
\begin{eqnarray}
|\vec{E}| > \left(\frac{2\pi^2}{\lambda}\right)^{1/2} m_q^2 \, .
\label{ecrmin}
\end{eqnarray}

\FIGURE[r]{ 
\includegraphics[width=7cm]{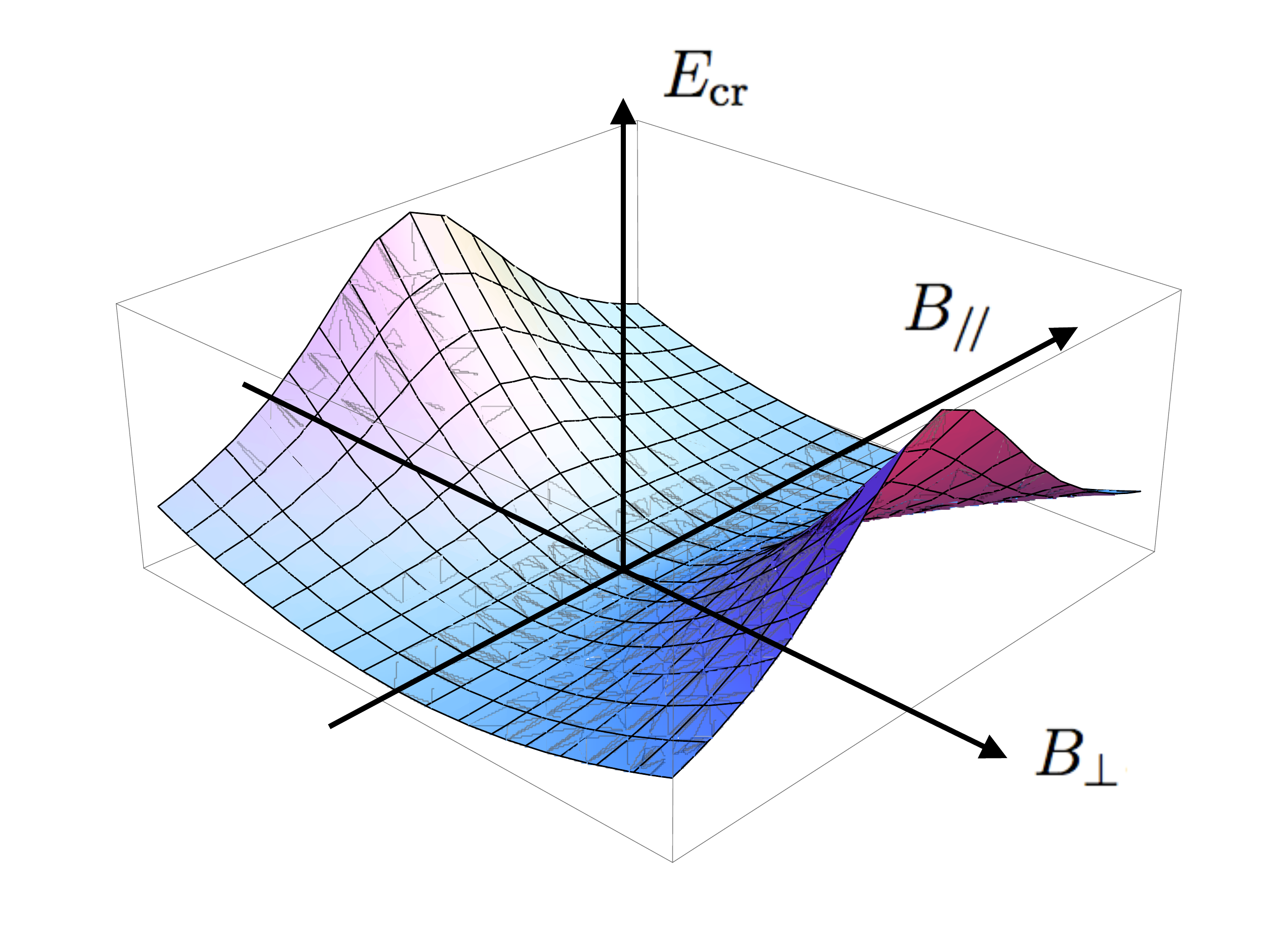}
\caption{A plot of the critical electric field $E_{\rm cr}$
as a function of the magnetic field $B_{/\!/}$
and $B_\perp$, for nonzero $m_q$. 
We find that the magnetic field 
makes the critical electric field larger.
}
\label{fig:ecr}
}
Let us study magnetic field dependence
of the critical electric field.
The critical electric field $E_{\rm cr}$ is a solution of the equation
\begin{eqnarray}
b(E_{\rm cr},B) = \left(\frac{2\pi^2}{\lambda}\right)^{1/4} m_q \, .
\label{Ecr}
\end{eqnarray}
Decomposing the magnetic field to two components $B_{/\!/}$ and $B_\perp$
(parallel / perpendicular to the electric field), we can plot the value of the critical electric field
as a function of the magnetic field $B_{/\!/}$ and $B_\perp$, see Fig.~\ref{fig:ecr}.
We find that the magnetic field does not lower the critical electric field.
In fact, the critical electric field is minimized when the perpendicular 
magnetic field $B_\perp$ 
vanishes, and the minimized value is equal to the critical electric field in
the absence of the magnetic field (\ref{ecrmin}). 
In the zero magnetic field case, the
critical field coincides with the confining force \cite{Hashimoto:2013mua}.
This is natural because the vacuum instability 
takes place when the quark-antiquarks are pulled apart 
with a force stronger than the confining force. 
It is strange that the critical field has a 
$B_\perp$ dependence because the gluons mediating the confinement force 
is not affected by the magnetic field. 
We leave this puzzle for future consideration.

\subsection{Vacuum decay rate in the small mass limit}
When the quark mass is finite, 
the vacuum decay rate, i.e., the imaginary part of the
effective action, is non-diverging, even 
above the critical field. 
Here, we evaluate the small mass asymptotic behavior of the 
vacuum decay rate. 
As shown in the previous section, the divergence of the 
decay rate originates  from the
integral at large $z$. 
With non-zero mass 
represented by the parameter $\eta \neq 0$, 
the function $\tilde{h}(z)$ has the following $z$ dependence ($z_* \equiv R^2/\eta$)
\begin{eqnarray}
\tilde{h}(z) \sim 
\left\{
\begin{array}{lc}
1 & (z \ll z_*), \\
\frac{\eta^2}{R^4}z^2 & (z \gg z_*).
\end{array}
\right. 
\end{eqnarray}
This alters the divergent behavior of the integral:
For $z \gg z_*$, the integrand of the effective Lagrangian
behaves as
\begin{eqnarray}
\frac{R^8}{z^5}\sqrt{\xi} \sim z^{-5} , 
\end{eqnarray}
whose integral is convergent. The leading behavior 
of the imaginar part is given by
\begin{eqnarray}
{\rm Im} \; {\cal L}
\sim 2\pi^2 \mu_7 \int_{z_0}^{z_*}
dz \; \frac{R^8}{z^5}
\frac{(2\pi\alpha')^2}{R^4} z^4 \left|\vec{E}\cdot \vec{B}\right|,
\label{div1}
\end{eqnarray}
where the massless limit corresponds to $z_*\rightarrow \infty$. 
The lower bound of the integral $z_0$ 
is determined by the condition that the Lagrangian becomes imaginary, 
and is the solution of equation
(\ref{z0def})
\begin{eqnarray}
z_0 \equiv \left(
\frac{2\pi\alpha'}{R^2} b^2 - \frac{\eta^2}{R^4}
\right)^{-1/2} \, .
\end{eqnarray}
The divergence appears when $z_0\ll z_*$, and the leading term is given by
(\ref{div1}) which is evaluated as
\begin{eqnarray}
{\rm Im} \; {\cal L}
&=&
\frac{N_c}{4\pi^2}
 \left|\vec{E}\cdot \vec{B}\right|
 \log \frac{z_*}{z_0} + \cdots
\nonumber 
\\
& = &
\frac{N_c}{4\pi^2}
 \left|\vec{E}\cdot \vec{B}\right|
  \log \frac{b(E,B)}{m_q}
 \; +\;  \mbox{higher in}\;\; \frac{m_q}{b(E,B)} \, .
 \label{regdiv}
\end{eqnarray}
Interestingly, this divergence coincides with (\ref{LT}) if we replace the temperature $T$ with the quark mass $m_q$. 
The linear relation between the temperature and the quark mass
is commonly found in thermal field theories at weak coupling, 
and our strongly coupled results are consistent with that.
We shall see later that our asymptotic behavior (\ref{regdiv}), 
agrees with a weak coupling calculation of ${\cal N}=2$ supersymmetric 
QED.

\subsection{Nonlinear optical response of the SQCD vacuum: Disappearance of the Cotton-Mouton effect}

The real part of the Euler-Heisenberg Lagrangian has
a meaning of electromagnetic vacuum polarization. In particular, 
if we expand it in terms of the electromagnetic fields, the coefficients
gives the non-linear optical response function. 
As for our system, the first non-trivial term is the 
fourth order term, which is related to the Kerr effect and
the Cotton-Mouton effect, i.e., birefringence induced by magnetic fields (see for a review, \cite{Battesti:2012hf}).
In the following, we study the non-linear optical response 
below the critical field by 
evaluating the
real part of the Euler-Heisenberg Lagrangian (\ref{xires}) as an expansion in terms 
of the electromagnetic fields.

Defining the Lorentz invariant combinations as
\begin{eqnarray}
F \equiv \vec{E}^2 - \vec{B}^2, \quad
G \equiv \vec{E}\cdot \vec{B}, 
\label{Lorentz}
\end{eqnarray}
our Euler-Heisenberg Lagrangian is written as
\begin{eqnarray}
{\cal L} = -2\pi^2 \mu_7 R^3 \int_0^\infty
\! dz \; \frac{R^5}{z^5}
\sqrt{1-g F - g^2 G^2}\, .
\label{fullEH}
\end{eqnarray}
Here we defined
\begin{eqnarray}
g \equiv \beta(z) \tilde{h}(z)^{-2} = \frac{(2\pi\alpha')^2}{R^4} z^4 \left(
1+\frac{\eta^2}{R^4}z^2\right)^{-2} \, .
\end{eqnarray}
If we expand the square root in (\ref{fullEH}) in terms of $F$ and $G^2$,
we notice that an integral of the form ($n (\geq 1)$)
\begin{eqnarray}
I_n \equiv \int_0^\infty dz \; \frac{R^5}{z^5} g(z)^n 
= R (2\pi\alpha')^2 \frac{1}{(1-n)(1-2n)}
\left(\frac{\lambda}{2\pi^2 m_q^4}\right)^{n-1} \, 
\end{eqnarray}
determines the expansion coefficients. 
The Euler-Heisenberg Lagrangian (\ref{fullEH}) is given by
\begin{eqnarray}
{\cal L} = -2\pi^2 \mu_7 R^3
\left[
I_0 - \frac12 F I_1 - \frac18 (F^2 + 4 G^2) I_2 + \cdots
\right].
\end{eqnarray}
The first term $I_0$ is divergent from the first place. It is the vacuum term
and we need an appropriate renormalization, as described in \cite{Hashimoto:2013mua}. 
The second term $I_1$ corresponds to the charge renormalization, since $F$ is
nothing but the original Maxwell electromagnetism Lagrangian.
Now, we come to the nontrivial leading correction $I_2$. 
Using the AdS/CFT dictionary for the coefficients, the term with $I_2$ is
expressed as
\begin{eqnarray}
{\cal L}_{\rm leading}
 = \frac{\lambda N_c}{3 \cdot 2^7 \pi^4 m_q^4} (4 G^2 + F^2) \, 
 \label{CMeff}
\end{eqnarray}
This is the leading electromagnetic correction to the effective action of strongly coupled 
${\cal N}=2$ supersymmetric QCD.

An interesting observation is that 
the Cotton-Mouton effect vanishes in
(\ref{CMeff}). The response function of Cotton-Mouton effect is given by a combination 
$c_{0,2}-4c_{2,0}$ where $c_{0,2} (c_{2,0})$ is the coefficient of $G^2 (F^2)$. We find that
\begin{eqnarray}
c_{0,2} = 4 \frac{\lambda N_c}{3 \cdot 2^7 \pi^4 m_q^4} = 4c_{2,0} \, ,
\end{eqnarray}
so the Cotton-Mouton effect vanishes. 

A possible reason for this result would be our supersymmetry (which is not present at low energy in nature), as opposed to the standard QED in which the Cotton-Mouton effect is non-vanishing, that is why experimental confirmation is expected. In fact, if we supersymmetrize the QED, the Cotton-Mouton effect vanishes, as we shall see in the next subsection.


\subsection{Coincidence with ${\cal N}=2$ supersymmetric QED}
In this subsection, we compare the 
Euler-Heisenberg Lagrangian of SQCD with 
the one-loop result of ${\cal N}=2$ supersymmetric QED.
A priori, we expect no relation 
between them because our SQCD is with self-interacting gluons and 
is evaluated at strong coupling through the AdS/CFT correspondence, while the SQED is a weak coupling and photons are not interacting with each other
at the one-loop level. 
However, unexpectedly, we find 
several coincidences: First is the
small electric field asymptotic of the vacuum decay rate, 
and the second is the leading nonlinear electromagnetic
response coefficients.
This agreement may be attributed to the supersymmetries.
It is known via AdS/CFT correspondence 
that in SQCD the gluon  has a Coulombic potential, that is presumably why 
we find the agreement in the following. 
So, in one aspect, 
our report here should serve as a consistency check of our calculation of the imaginary D-brane action in the AdS/CFT correspondence.

\subsubsection{Comparison of the vacuum decay rate}

First, let us check the divergence in the imaginary part. The one-loop QED \cite{Tanji:2008ku}\footnote{See \cite{Ritus:1978cj,Lebedev:1985bj,Affleck:1981bma} for calculations in 
QED. For supersymmetric calculations, see \cite{Buchbinder:1999jn,Kuzenko:2003qg}.}
has the following expression for the effective Lagrangian when $\vec{E}$ is parallel to $\vec{B}$;
\begin{eqnarray}
{\rm Im}\;{\cal L}_{\rm scalar} &=& \frac{EB}{8\pi^2}
\sum_{l=1}^{\infty} \frac{(-1)^{l+1}}{l} \frac{\exp[-\pi l m^2/E]}{2 \sinh (\pi l B/E)} \, ,
\\
{\rm Im}\;{\cal L}_{\rm spinor} &=& \frac{EB}{8\pi^2}
\sum_{l=1}^{\infty} \frac{1}{l} \exp[-\pi l m^2/E] \coth (\pi l B/E)  \, .
\end{eqnarray}
Here ${\cal L}_{\rm scalar}$ denotes scalar QED (the charged particle is a scalar bosonic field) 
and ${\cal L}_{\rm spinor}$ is for
the ordinary QED. To have ${\cal N}=2$ supersymmetry, we need $2N_c$ scalars
and $N_c$ spinors, and thus
\begin{eqnarray}
{\rm Im} \; {\cal L}_{{\cal N}=2 \; {\rm SQED}}
&=& N_c \left(
{\rm Im}\;{\cal L}_{\rm spinor}  + 2 \; {\rm Im}\;{\cal L}_{\rm scalar} 
\right)
\nonumber \\
&=& 
\frac{N_c EB}{8\pi^2}
\sum_{l=1}^{\infty} \frac{1}{l} \exp[-\pi l m^2/E] 
\frac{\cosh (\pi l B/E) + (-1)^{l+1}}{\sinh (\pi l B/E)}  \, .
\end{eqnarray}
Let us consider a limit of electron mass $m$ going to zero. In the expression above,
the factor $\exp[-\pi l m^2/E] $ serves as a  cut-off of the summation over $l$. Therefore
we can approximate it as
\begin{eqnarray}
{\rm Im} \; {\cal L}_{{\cal N}=2 \; {\rm SQED}} \sim 
\frac{N_c EB}{8\pi^2}
\sum_{l=1}^{E/\pi m^2} \frac{1}{l} 
\frac{\cosh (\pi l B/E) + (-1)^{l+1}}{\sinh (\pi l B/E)}  \, .
\end{eqnarray}
The divergence is due to $\cosh/\sinh \sim 1$ for large $l$, so, for large $E/m^2$ we can
further approximate it as\footnote{
In literature, this expression for the dominant imaginary part in QED (non-supersymmetric) can be found in \cite{Hidaka:2011fa,Hidaka:2011dp}.
}
\begin{eqnarray}
{\rm Im} \; {\cal L}_{{\cal N}=2 \; {\rm SQED}} &\sim &
\frac{N_c EB}{8\pi^2}
\sum_{l=1}^{E/\pi m^2} \frac{1}{l} 
\sim \frac{N_c EB}{8\pi^2}
\log \frac{E}{\pi m^2} 
\nonumber \\
&\sim &
\frac{N_c}{4\pi^2}EB
\log \frac{\sqrt{E}}{m}  \, .
\end{eqnarray}
We find that this SQED result is in  agreement with our SQCD result (\ref{regdiv}).

\subsubsection{Comparison of the nonlinear optical response coefficient}

Next, we look at the real part when the electromagnetic field is small. 
We shall see that, again, our SQCD result coincides with that of the SQED.

The integral expression for the Euler-Heisenberg Lagrangian
for spinor and scalar QED (for a review, see \cite{Dunne:2004nc}) is given as
\begin{eqnarray}
{\cal L}_{\rm scalar}
&=& \frac{1}{16 \pi^2}
\int_0^\infty \! d\eta \; \frac{e^{-\eta m_e^2}}{\eta^3}
\left[
\frac{ab\eta^2}{\sinh b\eta \sin a\eta}-1 + \frac{\eta^2}{6}(b^2-a^2)
\right] \, ,
\label{sqed}
\\
{\cal L}_{\rm spinor}
&=& \frac{-1}{8 \pi^2}
\int_0^\infty \! d\eta \; \frac{e^{-\eta m_e^2}}{\eta^3}
\left[
\frac{ab\eta^2}{\tanh b\eta \tan a\eta}-1 - \frac{\eta^2}{3}(b^2-a^2)
\right] \, .
\label{qed}
\end{eqnarray}
Here the real constants $a$ and $b$ are are related to the 
electro-magnetic fields via
$a^2-b^2 = \vec{E}^2-\vec{B}^2$ and $ab = \vec{E}\cdot\vec{B}$. We can combine the two expressions to obtain the Euler-Heisenberg Lagrangian for the
${\cal N}=2$ supersymmetric QED 
\begin{eqnarray}
&&  {\cal L}_{{\cal N}=2 \; {\rm SQED}}
= N_c \left(
{\cal L}_{\rm spinor}  + 2 \; {\cal L}_{\rm scalar} \right)
\nonumber \\
&& =  
\frac{-N_c}{8 \pi^2}
\int_0^\infty \! d\eta \; \frac{e^{-\eta m_e^2}}{\eta^3}
\left[
\frac{ab\eta^2(\cosh b \eta \cos a \eta -1)}{\sinh b\eta \sin a\eta} - \frac{\eta^2}{6}(b^2-a^2)
\right] \, .
\end{eqnarray}
The constant term in the integral of (\ref{sqed}) and (\ref{qed}) 
correspond to vacuum energy, and cancel each other
due to the supersymmetries. We can expand this expression for small $a$ and $b$ 
\begin{eqnarray}
{\cal L}_{{\cal N}=2 \; {\rm SQED}}
=\frac{-N_c}{8 \pi^2}
\int_0^\infty \! d\eta \; \frac{e^{-\eta m_e^2}}{\eta}
\left[
-\frac13(a^2-b^2)
- \eta^2
 \frac{(a^2+b^2)^2}{24} + {\cal O}((a,b)^6)
 \right] \, .
\end{eqnarray}
The first term represents charge renormalization and divergent, so here we ignore it.
We are interested in the second term, i.e., the leading 
nontrivial correction to SQED. After the integration over $\eta$, we obtain
\begin{eqnarray}
{\cal L}^{{\cal N}=2 \; {\rm SQED}}_{\rm leading}
=
\frac{N_c}{2^6\cdot 3 \pi^2 m_e^4}(a^2+b^2)^2
=
\frac{N_c}{2^6\cdot 3 \pi^2 m_e^4}\left(4G^2+F^2\right)
\end{eqnarray}
where $F$ and $G$ are defined in (\ref{Lorentz}). 

Surprisingly, 
we find that this SQED result coincides
with the SQCD result (\ref{CMeff}).
This is obtained if we substitute the following relation
\begin{eqnarray}
m_e^2 \leftrightarrow \frac{\sqrt{2}\pi}{\sqrt{\lambda}} m_q^2 \, .
\label{relation}
\end{eqnarray}
The value $\frac{\sqrt{2}\pi}{\sqrt{\lambda}} m_q^2$ is equal to $E_{\rm cr}$ which
we was found in \cite{Hashimoto:2013mua}, and this relation was identical to that
found in \cite{Hashimoto:2013mua} to find an agreement between the imaginary parts
of SQED and SQCD for large electric fields. 

In summary, in this subsection, we found agreement in asymptotic behaviors
of the vacuum decay rate and the nonlinear optical response coefficients
between SQCD at the strong coupling and with non-interacting SQED (one loop),
assuming a natural relation (\ref{relation}).

\section{Summary}

In this work, we studied the response of the vacuum
in strong electric and magnetic fields 
by calculating the Euler-Heisenberg Lagrangian (\ref{xires}) of 
the strongly coupled ${\cal N}=2$ supersymmetric QCD 
in the large $N_c$ limit. 
Above the critical electric field, the vacuum becomes
unstable against quark-anti-quark pair production 
that takes place due to the Schwinger mechanism. 
The vacuum decay rate is given by the 
imaginary part of the Euler-Heisenberg Lagrangian.
The real part is the generating function of 
response functions of non-linear optical processes. 
The AdS/CFT correspondence enables us to evaluate 
the Euler-Heisenberg Lagrangian. 

We find that the imaginary part of the effective action diverges when the quark mass
(in this theory it is roughly equal to the confinement scale) approaches zero. 
This divergence appears only in the presence of magnetic fields
and we attribute this to the effective dimensional reduction
due to Landau quantization of the quarks.
We calculated the temperature and quark mass dependence of the decay rate 
in section 2 and 3 respectively. 
We found out that the critical electric field
depends on the magnetic field  (\ref{limitE}). 
In our ${\cal N}=2$ SQCD, the magnetic field makes 
the critical electric field larger. We do not fully understand 
the reason of this magnetic field dependence.

We also found, in some limits,
the Euler-Heisenberg Lagrangian of large $N_c$ SQCD
in the strong coupling limit 
agrees with non-interacting ${\cal N}=2$ supersymmetric QED. 
The imaginary part at strong electromagnetic field, and the other is the real part in the weak field expansion, at fourth order.
This is an interesting coincidence which 
we did not expect: The background electromagnetic field
breaks the supersymmetry so this coincidence cannot be explained by it. 
Although we do not understand this, our
finding may serve as an evidence that the AdS/CFT
dictionary can be extended to the false vacuum. This encourages us to 
further investigate non-supersymmetric setup in the AdS/CFT correspondence
at which we can evaluate the Euler-Heisenberg Lagrangian and the Schwinger effect
at more realistic strongly coupled field theories. We shall report this elsewhere \cite{future}.

\section*{Acknowledgments}

K.H. would like to thank Y.~Hidaka.  We are grateful to the
hospitality of OIST. 
K.H. and 
T.O. are supported by KAKENHI (Grant No. 23654096, 23740260, 24224009).
This research was
partially supported by the RIKEN iTHES project.


\include{end}

\begin{thebibliography}{99}
\bibitem{Schwinger:1951} 
  J.~S.~Schwinger,
  ``On gauge invariance and vacuum polarization,''
  Phys.\ Rev.\  {\bf 82}, 664 (1951).

\bibitem{Dunne:2004nc} 
  G.~V.~Dunne,
  ``Heisenberg-Euler effective Lagrangians: Basics and extensions,''
  In *Shifman, M. (ed.) et al.: From fields to strings, vol. 1* 445-522
  [hep-th/0406216].

\bibitem{Dittrich}
W. Dittrich and H. Gies, 
``Probing the Quantum Vacuum."  Springer-Verlag, Berlin, 2000.

\bibitem{Heisenberg:1935qt} 
  W.~Heisenberg and H.~Euler,
  ``Consequences of Dirac's theory of positrons,''
  Z.\ Phys.\  {\bf 98}, 714 (1936)



\bibitem{Hashimoto:2013mua} 
  K.~Hashimoto and T.~Oka,
  ``Vacuum Instability in Electric Fields via AdS/CFT: Euler-Heisenberg Lagrangian and Planckian Thermalization,''
  JHEP {\bf 1310}, 116 (2013)
  [arXiv:1307.7423].
  
\bibitem{Maldacena:1997re} 
  J.~M.~Maldacena,
  ``The Large N limit of superconformal field theories and supergravity,''
  Adv.\ Theor.\ Math.\ Phys.\  {\bf 2}, 231 (1998)
  [hep-th/9711200].

\bibitem{Gubser:1998bc}
 S.~Gubser, I.~R. Klebanov, and A.~M. Polyakov, 
 ``Gauge theory correlators from noncritical string theory,"  {\em Phys.Lett.} B {\bf 428},105 (1998).
 
 \bibitem{Witten:1998qj}
 E.~Witten, 
 ``Anti-de Sitter space and holography,"  {\em
   Adv.Theor.Math.Phys.} {\bf 2}, 253 (1998).
 
\bibitem{Yildiz:1979vv} 
  A.~Yildiz and P.~H.~Cox,
  ``Vacuum Behavior in Quantum Chromodynamics,''
  Phys.\ Rev.\ D {\bf 21}, 1095 (1980).

\bibitem{Ambjorn:1982bp} 
  J.~Ambjorn and R.~J.~Hughes,
  ``Canonical Quantization in Nonabelian Background Fields. 1.,''
  Annals Phys.\  {\bf 145}, 340 (1983).
  
\bibitem{Tanji:2008ku} 
  N.~Tanji,
  ``Dynamical view of pair creation in uniform electric and magnetic fields,''
  Annals Phys.\  {\bf 324}, 1691 (2009)
  [arXiv:0810.4429 [hep-ph]].
  
\bibitem{Tanji:2011di} 
  N.~Tanji and K.~Itakura,
  ``Schwinger mechanism enhanced by the Nielsen--Olesen instability,''
  Phys.\ Lett.\ B {\bf 713}, 117 (2012)
  [arXiv:1111.6772 [hep-ph]].



\bibitem{Kharzeev:2007jp} 
  D.~E.~Kharzeev, L.~D.~McLerran and H.~J.~Warringa,
  ``The Effects of topological charge change in heavy ion collisions: 'Event by event P and CP violation',''
  Nucl.\ Phys.\ A {\bf 803}, 227 (2008)
  [arXiv:0711.0950 [hep-ph]].
  
\bibitem{Skokov:2009qp} 
  V.~Skokov, A.~Y.~.Illarionov and V.~Toneev,
  ``Estimate of the magnetic field strength in heavy-ion collisions,''
  Int.\ J.\ Mod.\ Phys.\ A {\bf 24}, 5925 (2009)
  [arXiv:0907.1396 [nucl-th]].

\bibitem{Voronyuk:2011jd} 
  V.~Voronyuk, V.~D.~Toneev, W.~Cassing, E.~L.~Bratkovskaya, V.~P.~Konchakovski and S.~A.~Voloshin,
  ``(Electro-)Magnetic field evolution in relativistic heavy-ion collisions,''
  Phys.\ Rev.\ C {\bf 83}, 054911 (2011)
  [arXiv:1103.4239 [nucl-th]].
  
\bibitem{Bzdak:2011yy} 
  A.~Bzdak and V.~Skokov,
  ``Event-by-event fluctuations of magnetic and electric fields in heavy ion collisions,''
  Phys.\ Lett.\ B {\bf 710}, 171 (2012)
  [arXiv:1111.1949 [hep-ph]].


\bibitem{Deng:2012pc} 
  W.~-T.~Deng and X.~-G.~Huang,
  ``Event-by-event generation of electromagnetic fields in heavy-ion collisions,''
  Phys.\ Rev.\ C {\bf 85}, 044907 (2012)
  [arXiv:1201.5108 [nucl-th]].

\bibitem{Semenoff:2011ng} 
  G.~W.~Semenoff and K.~Zarembo,
  ``Holographic Schwinger Effect,''
  Phys.\ Rev.\ Lett.\  {\bf 107}, 171601 (2011)
  [arXiv:1109.2920 [hep-th]].
  
\bibitem{Ambjorn:2011wz} 
  J.~Ambjorn and Y.~Makeenko,
  ``Remarks on Holographic Wilson Loops and the Schwinger Effect,''
  Phys.\ Rev.\ D {\bf 85}, 061901 (2012)
  [arXiv:1112.5606 [hep-th]].

\bibitem{Bolognesi:2012gr} 
  S.~Bolognesi, F.~Kiefer and E.~Rabinovici,
  ``Comments on Critical Electric and Magnetic Fields from Holography,''
  JHEP {\bf 1301}, 174 (2013)
  [arXiv:1210.4170 [hep-th]].

\bibitem{Sato:2013pxa} 
  Y.~Sato and K.~Yoshida,
  ``Holographic description of the Schwinger effect in electric and magnetic fields,''
  JHEP {\bf 1304}, 111 (2013)
  [arXiv:1303.0112 [hep-th]].

\bibitem{Sato:2013iua} 
  Y.~Sato and K.~Yoshida,
  ``Potential Analysis in Holographic Schwinger Effect,''
  JHEP {\bf 1308}, 002 (2013)
  [arXiv:1304.7917, arXiv:1304.7917 [hep-th]].
  
\bibitem{Sato:2013dwa} 
  Y.~Sato and K.~Yoshida,
  ``Holographic Schwinger effect in confining phase,''
  JHEP {\bf 1309}, 134 (2013)
  [arXiv:1306.5512 [hep-th]].

\bibitem{Sato:2013hyw} 
  Y.~Sato and K.~Yoshida,
  ``Universal aspects of holographic Schwinger effect in general backgrounds,''
  JHEP {\bf 1312}, 051 (2013)
  [arXiv:1309.4629 [hep-th]].

\bibitem{Kawai:2013xya} 
  D.~Kawai, Y.~Sato and K.~Yoshida,
  ``The Schwinger pair production rate in confining theories via holography,''
  arXiv:1312.4341 [hep-th].

\bibitem{Sakaguchi:2014gpa} 
  M.~Sakaguchi, H.~Shin and K.~Yoshida,
  ``No pair production of open strings in a plane-wave background,''
  arXiv:1402.2048 [hep-th].


\bibitem{Gorsky:2001up} 
  A.~S.~Gorsky, K.~A.~Saraikin and K.~G.~Selivanov,
  ``Schwinger type processes via branes and their gravity duals,''
  Nucl.\ Phys.\ B {\bf 628}, 270 (2002)
  [hep-th/0110178].

\bibitem{Sonner:2013mba} 
  J.~Sonner,
  ``Holographic Schwinger Effect and the Geometry of Entanglement,''
  Phys.\ Rev.\ Lett.\  {\bf 111}, 211603 (2013)
  [arXiv:1307.6850 [hep-th]].

\bibitem{Chernicoff:2013iga} 
  M.~Chernicoff, A.~G\"uijosa and J.~F.~Pedraza,
  ``Holographic EPR Pairs, Wormholes and Radiation,''
  JHEP {\bf 1310}, 211 (2013)
  [arXiv:1308.3695 [hep-th]].

\bibitem{Karch}
A.~Karch and E.~Katz, 
``Adding flavor to AdS / CFT," 
  JHEP {\bf 0206} 043 (2002), [arXiv:hep-th/0205236].

\bibitem{Karch2}
A.~Karch and A.~O'Bannon, 
``Metallic AdS/CFT," 
JHEP {\bf 0709}, 024 (2007)
 [arXiv:0705.3870 [hep-th]].

\bibitem{Karch3}
A.~Karch and A.~O'Bannon, 
 ``Holographic thermodynamics at finite baryon density: Some exact results," 
JHEP {\bf 0711}, 074 (2007) 
 [arXiv:0709.0570 [hep-th]].

\bibitem{Erdmenger}
J.~Erdmenger, R.~Meyer, and J.~P.~Shock, 
 ``AdS/CFT with flavour in electric and magnetic Kalb-Ramond fields," 
JHEP 0712, 091 (2007)
 [arXiv:0709.1551 [hep-th]].




\bibitem{Hartnoll:2007ip} 
  S.~A.~Hartnoll and C.~P.~Herzog,
  ``Ohm's Law at strong coupling: S duality and the cyclotron resonance,''
  Phys.\ Rev.\ D {\bf 76}, 106012 (2007)
  [arXiv:0706.3228 [hep-th]].
  
\bibitem{Albash:2007bk} 
  T.~Albash, V.~G.~Filev, C.~V.~Johnson and A.~Kundu,
  ``Finite temperature large N gauge theory with quarks in an external magnetic field,''
  JHEP {\bf 0807}, 080 (2008)
  [arXiv:0709.1547 [hep-th]].
  

\bibitem{Albash:2007bq} 
  T.~Albash, V.~G.~Filev, C.~V.~Johnson and A.~Kundu,
  ``Quarks in an external electric field in finite temperature large N gauge theory,''
  JHEP {\bf 0808}, 092 (2008)
  [arXiv:0709.1554 [hep-th]].

\bibitem{Johnson:2008vna} 
  C.~V.~Johnson and A.~Kundu,
  ``External Fields and Chiral Symmetry Breaking in the Sakai-Sugimoto Model,''
  JHEP {\bf 0812}, 053 (2008)
  [arXiv:0803.0038 [hep-th]].

\bibitem{Bergman:2008sg} 
  O.~Bergman, G.~Lifschytz and M.~Lippert,
  ``Response of Holographic QCD to Electric and Magnetic Fields,''
  JHEP {\bf 0805}, 007 (2008)
  [arXiv:0802.3720 [hep-th]].

\bibitem{Kim:2008zn} 
  K.~-Y.~Kim, S.~-J.~Sin and I.~Zahed,
  ``Dense and Hot Holographic QCD: Finite Baryonic E Field,''
  JHEP {\bf 0807}, 096 (2008)
  [arXiv:0803.0318 [hep-th]].

\bibitem{Evans:2011mu} 
  N.~Evans, A.~Gebauer and K.~-Y.~Kim,
  ``$E, B, \mu, T$ Phase Structure of the D3/D7 Holographic Dual,''
  JHEP {\bf 1105}, 067 (2011)
  [arXiv:1103.5627 [hep-th]].

\bibitem{Alam:2012fw} 
  M.~S.~Alam, V.~S.~Kaplunovsky and A.~Kundu,
  ``Chiral Symmetry Breaking and External Fields in the Kuperstein-Sonnenschein Model,''
  JHEP {\bf 1204}, 111 (2012)
  [arXiv:1202.3488 [hep-th]].
  
  
  
\bibitem{Bigazzi:2013jqa} 
  F.~Bigazzi, A.~L.~Cotrone and J.~Tarrio,
  ``Charged D3-D7 plasmas: novel solutions, extremality and stability issues,''
  JHEP {\bf 1307}, 074 (2013)
  [arXiv:1304.4802 [hep-th]].



\bibitem{Battesti:2012hf} 
  R.~Battesti and C.~Rizzo,
  ``Magnetic and electric properties of quantum vacuum,''
  Rept.\ Prog.\ Phys.\  {\bf 76}, 016401 (2013)
  [arXiv:1211.1933 [physics.optics]].




\bibitem{Ritus:1978cj} 
  V.~I.~Ritus,
  ``Method Of Eigenfunctions And Mass Operator In Quantum Electrodynamics Of A Constant Field,''
  Sov.\ Phys.\ JETP {\bf 48}, 788 (1978)
  [Zh.\ Eksp.\ Teor.\ Fiz.\  {\bf 75}, 1560 (1978)].

\bibitem{Lebedev:1985bj} 
  S.~L.~Lebedev and V.~I.~Ritus,
  ``Virial Representation Of The Imaginary Part Of The Lagrange Function Of The Electromagnetic Field,''
  Sov.\ Phys.\ JETP {\bf 59}, 237 (1984)
  [Zh.\ Eksp.\ Teor.\ Fiz.\  {\bf 86}, 408 (1984)].

\bibitem{Affleck:1981bma} 
  I.~K.~Affleck, O.~Alvarez and N.~S.~Manton,
  ``Pair Production at Strong Coupling in Weak External Fields,''
  Nucl.\ Phys.\ B {\bf 197}, 509 (1982).

\bibitem{Buchbinder:1999jn} 
  I.~L.~Buchbinder, S.~M.~Kuzenko and A.~A.~Tseytlin,
  ``On low-energy effective actions in N=2, N=4 superconformal theories in four-dimensions,''
  Phys.\ Rev.\ D {\bf 62}, 045001 (2000)
  [hep-th/9911221].

\bibitem{Kuzenko:2003qg} 
  S.~M.~Kuzenko and I.~N.~McArthur,
  ``Low-energy dynamics in N=2 super QED: Two loop approximation,''
  JHEP {\bf 0310}, 029 (2003)
  [hep-th/0308136].



\bibitem{Hidaka:2011fa} 
  Y.~Hidaka, T.~Iritani and H.~Suganuma,
  ``Fast Vacuum Decay into Quark Pairs in Strong Color Electric and Magnetic Fields,''
  AIP Conf.\ Proc.\  {\bf 1388}, 516 (2011)
  [arXiv:1103.3097 [hep-ph]].

\bibitem{Hidaka:2011dp} 
  Y.~Hidaka, T.~Iritani and H.~Suganuma,
  ``Fast vacuum decay into particle pairs in strong electric and magnetic fields,''
  arXiv:1102.0050 [hep-ph].

  
\bibitem{future}
  K.~Hashimoto, T.~Oka and A.~Sonoda, work in progress.
  \end{thebibliography}
\end{document}